\documentclass[fleqn,a4paper,12pt]{article}
\usepackage{amssymb}
\usepackage{amsmath}
\usepackage[dvips]{graphicx}
\usepackage{placeins}
\usepackage{lscape}
\usepackage{a4}
\usepackage{epsfig}

\newcommand{\ri}{{\mathrm i}}
\newcommand{\p}{\partial}
\newcommand{\bea}{\begin{array}}
\newcommand{\eea}{\end{array}}

\catcode`@=11 \long
\def\@caption#1[#2]#3{\par\addcontentsline{\csname
ext@#1\endcsname}{#1} {\protect\numberline{\csname
the#1\endcsname}{\ignorespaces #2}} \begingroup \small
\@parboxrestore \@makecaption{\csname fnum@#1\endcsname}
{\ignorespaces #3}\par \endgroup} \catcode`@=12

\newcommand{\la}{\label}

\catcode`@=11 \long
\def\@caption#1[#2]#3{\par\addcontentsline{\csname
ext@#1\endcsname}{#1} {\protect\numberline{\csname
the#1\endcsname}{\ignorespaces #2}} \begingroup \small
\@parboxrestore \@makecaption{\csname fnum@#1\endcsname}
{\ignorespaces #3}\par \endgroup} \catcode`@=12

\begin{document}

\allowdisplaybreaks
 \begin{titlepage} \vskip 2cm

\begin{center} {\Large\bf Laplace-Runge-Lenz vector with spin in any dimension}
\footnote{E-mail:
{\tt nikitin@imath.kiev.ua} } \vskip 3cm {\bf {A. G. Nikitin }
\vskip 5pt {\sl Institute of Mathematics, National Academy of
Sciences of Ukraine,\\ 3 Tereshchenkivs'ka Street, Kyiv-4, Ukraine,
01601\\}}\end{center}
\vskip .5cm \rm
\begin{abstract} Superintegrable $d$ - dimensional quantum mechanical systems with spin, which admit a generalized Laplace-Runge-Lenz  vector are presented. The systems with spins 0, $\frac12$ and 1 are considered in detail. All these systems are exactly solvable for arbitrary $d$, and their solutions are presented explicitly.

\end{abstract}
\end{titlepage}
\section{Introduction\label{int}}
Exactly solvable systems  present perfect tools for understanding of  grounds of quantum mechanics. A very attractive property of the majority of such systems is that they admit extended  symmetries which are nice and important signs of their solvability.

Maybe the most important exactly solvable system is the Hydrogen atom. In addition to transparent invariance with respect to the rotation group SO(3), this system admits a hidden (Fock) symmetry with respect to group SO(4) \cite{Fock}. In other words, the Hydrogen atom is a superintegrable system. In addition, it admits a supersymmetry, since its potential is shape invariant with respect to a special Darboux transform \cite{khare}.

There are numerous reasons to search for other superintegrable and supersymmetric quantum mechanical systems. In particular, that is a promissing way to discover new exactly solvable systems. Integrable and superintegrable systems for $d=2$ have been first classified in papers \cite{wint1} and \cite{BM}. 3d systems with second order integrals of motion are described in (\cite{ev1}) and (\cite{ev2}). A fresh survey of the classic and contemporary results in this field can be found in \cite{wint01}.

Supersymmetric systems with scalar potentials belong to a well developed and in some aspects almost completed research field, see survey \cite{khare}. However, the recent discovery of an infinite family of new shape invariant systems generated by the exceptional polynomials \cite{quesne} opens new interesting ways in this area.

A much less studied field includes shape invariant systems with matrix potentials. A systematic search for such systems was started with papers \cite{N3} and \cite{N4}, also particular examples of shape invariant matrix potentials where discussed in many papers, see, e.g., \cite{fu} and \cite{nb}. The completed list of additive shape invariant $2\times2$ matrix potentials is presented in paper \cite{N4}.

Matrix potentials appears naturally in all problems including quantum mechanical particles with spin.  Superintegrable systems with spin which include the spin-orbit interaction were studied in papers \cite{w6}, \cite{w7} and  \cite{w8}.

But there is one more type of spin interaction which has perfect grounds in quantum mechanics. It is the Pauli interaction which can be represented, e.g., by the Stern-Gerlach term $~{\bf S}\cdot{\bf B}$ where $\bf S$ and ${\bf B}$  are a spin vector and a vector of magnetic field strength correspondingly.

A perfect example of an exactly solvable system with Pauli interaction, which is both supersymmetric and superintegrable, was discovered some time ago by Pron'ko and Stroganov
\cite{Pron}. This system is effectively planar and simulates a neutral particle with a non-trivial magnetic or electric dipole momentum (e.g., neutron). Like the Hydrogen atom, the Pron'ko-Stroganov (PS) system admits a vector  integral of motion which is a $2d$ quantum analogue of the Laplace-Runge-Lenz (LRL) vector. Moreover, this system describes a particle with spin $\frac12$. A relativistic analogue of the PS system was obtained in \cite{N10}, the other new
2d and 3d superintegrable systems with spin $\frac12$ were discussed in \cite{N11}.

 Only over thirty years a generalization of the PS system to the case of arbitrary spin was obtained \cite{Pron2}. Like the initial PS model, the systems with arbitrary spin are superintegrable and  supersymmetric. They have been integrated in \cite{N5} using shape invariant superpotentials classified in \cite{N3}.

 A new exactly solvable $3d$ system with Fock symmetry was discussed in \cite{N1}. That is a generalization of the PS system to the tree-dimensional case. The systems proposed in \cite{N1} are supersymmetric and admits a hidden symmetry with respect to group $SO(4)$. In other words, they keep all symmetries admitted by the Hydrogen atom, and so can be solved exactly \cite{N1}. But, in contrast with the non-relativistic Hydrogen atom, these systems include  orbital  particles with spin $\frac12$.

 The next natural step was to search for $3d$ systems with arbitrary spin, which keep the hidden symmetry with respect to group $SO(4)$. Such systems were presented in paper \cite{N6}. In this way a generalization of the LRL vector for systems including orbital particles with arbitrary spin has been constructed.

 In the present paper the discussion of superintegrable systems with spin  is extended to the case of arbitrary dimensional space. More exactly, the $d$ dimensional systems which admit a LRL vector are presented. The systems with spin 0, $\frac12$ and 1 are considered in detail. All these systems appears to be exactly solvable for arbitrary $d$, and their solutions are presented explicitly.

\section{ Schr\"odinger equations in d-dimensional space }
\subsection{General analysis}
Consider a $d$-dimensional stationary Schr\"odinger equation with a matrix potential
\begin{gather}H\Psi\equiv\left(\frac{p^2}{2m} +V({\bf x})\right)\Psi=E\Psi\la{se}\end{gather}
where $p^2=p_1^2+p_2^2+...+p_d^2$, $p_1=-\ri\frac{\p}{\p x_1}$, and $V$ is a matrix potential dependent on ${\bf x}=\left(x_1, x_2, ..., x_d\right)$.

We suppose Hamiltonian $H$ be invariant with respect to the rotation group SO($d$) whose generators can be chosen in  the standard form:
\begin{gather}J_{\mu\nu}=x_\mu p_\nu-x_\nu p_\mu+S_{\mu\nu}\la{Jmn}\end{gather}
where indices $\mu$ and $\nu$ run over the values $1, 2, ..., d$, and   $S_{\mu\nu}$ are matrices satisfying the familiar so($d$) commutation relations:
\begin{gather}\la{Smn}[S_{\mu\nu},S_{\lambda\sigma}]=
\ri(\delta_{\mu\lambda}S_{\nu\sigma}+\delta_{\nu\sigma}S_{\mu\lambda}-
\delta_{\mu\sigma}S_{\nu\lambda}-\delta_{\nu\lambda}
S_{\mu\sigma})\end{gather}
where $\delta_{\mu\lambda}$ is the Kronecker symbol.
By definition the matrix potential $V({\bf x})$ should commute with generators (\ref{Jmn}):
\begin{gather}[ V, J_{\mu\nu}]=0.\la{c1}\end{gather}

Let us search for such equations (\ref{se}) which admit additional integrals of motion $K_\mu, \ \mu=1,2,...,d$ of the following generic form:
\begin{gather}\la{lrl}K_\mu=\frac1{2m}\left(p_\nu J_{\mu\nu}+
J_{\mu\nu}p_\nu\right)+x_\mu V.\end{gather}
By definition,  $K_\mu$ should commute with $H$. This condition generates the following equations for potential:
\begin{gather}
{ x_\nu}\nabla_\nu  V+V=0,\la{c2}\\
 S_{\mu\nu}\nabla_\nu V+\nabla_\nu VS_{\mu\nu}
=0\la{c3}\end{gather} where $\nabla_\nu=\frac{\p}{\p { x_\nu}}$
and summation from 1 to $d$ is imposed over the repeating index $\nu$.

Equations (\ref{c1}), (\ref{c2}) and (\ref{c3}) present the necessary and sufficient conditions of the commutativity of hamiltonian  $H$ with operators $J_{\mu\nu}$ and $K_\nu$. For $d=2$ and $d=3$ these equations can be reduced to the commutativity conditions obtained in \cite{Pron2} and \cite{N1} for two- and three-dimensional systems respectively.

If conditions (\ref{c1}), (\ref{c2}) and (\ref{c3}) are fulfilled then operators $J_{\mu\nu}$ and $K_\mu$ satisfy the following relations:
\begin{gather}\la{a2}[J_{\mu\nu}, H]=[ K_\mu, H]=0,
\\\la{a3}\begin{split}&[ K_\mu,J_{\nu\lambda}]=i(\delta_{\mu\lambda} K_\nu-\delta_{\mu\nu}\hat K_\lambda),\\&[ K_\mu, K_\nu]=-\frac{2\ri}mJ_{\mu\nu}H,\end{split}\\
 [J_{\mu\nu},J_{\lambda\sigma}]=\ri(\delta_{\mu\lambda}J_{\nu\sigma}+
 \delta_{\nu\sigma}J_{\mu\lambda}-
\delta_{\mu\sigma}J_{\nu\lambda}-\delta_{\nu\lambda}J_{\mu\sigma}).\la{a4}\end{gather}

If $ H$ is changed by its eigenvalue $E$  then relations (\ref{a3}) and (\ref{a4})  define a Lie  algebra isomorphic to so($d+1$) if $E<0$ and to so($1,d$) for $E$ positive. In the special case $E=0$ we obtain the Lie algebra of the Euclidean group in $d$ dimensions. More exactly, for $E\neq0$ operators $J_{\mu\nu}$
and \begin{gather}\la{JJ}J_{d+1 \nu}=\sqrt{-\frac{m}{2E}}{ K}_\nu\end{gather} satisfy the following commutation relations:
\begin{gather*}\la{a44}[J_{\mu\nu},J_{\lambda\sigma}]=\ri(g_{\mu\lambda}J_{\nu\sigma}+
 g_{\nu\sigma}J_{\mu\lambda}-
g_{\mu\sigma}J_{\nu\lambda}-g_{\nu\lambda}J_{\mu\sigma})\end{gather*}
where all subindices run over values 1, 2, ..., $d+1$ and non-zero entries of tensor  $g_{\mu\nu}$ are
$$g_{11}=g_{22}=...=g_{dd}=-\text{sign}(E)g_{d+1\ d+1}=-1.$$

In other words, all systems whose potentials satisfy conditions (\ref{c1}), (\ref{c2}) and (\ref{c3}) admit a hidden symmetry of Fock type. The corresponding integral of motion (\ref{lrl}) is an analogue of the LRL vector for $d-$dimensional space.

This hidden symmetry makes it possible to impose an additional condition on $\Psi$, which fixes eigenvalue $\omega$ of the second order Casimir operator $C_2=\frac12J_{\mu\nu}J_{\mu\nu}$:
\begin{gather}\la{acon}\frac12J_{\mu\nu}J_{\mu\nu}\Psi=\omega\Psi, \quad \mu, \nu=1,2,...d+1\end{gather}
 Using (\ref{Jmn}), (\ref{lrl}) and (\ref{c2}) this condition can be rewritten in the following form:
 \begin{gather}\la{AK}\begin{split}&\left(-\frac12S_{\mu\nu}S_{\mu\nu}p^2+
 S_{\lambda\mu}S_{\lambda\nu}p_\mu p_\nu-\frac{ m}2\left(S_{\mu\nu}J_{\mu\nu}V+VS_{\mu\nu}J_{\mu\nu}
 \right)+ m^2r^2V^2\right)\Psi\\&=-2mE\left(\frac{(d-1)^2}4+\omega\right)\Psi\end{split}\end{gather}

 Thus equation (\ref{se}) can be supplemented by additional equation (\ref{AK}) which should be compatible with (\ref{se}). We will see that the compatibility condition for this system presents effective tools for solving the initial equation (\ref{se}).
\subsection{Scalar systems}

Let us start with a very particular case when matrices $S_{\mu\nu}$ in (\ref{Jmn}) are trivial. The corresponding system (\ref{se}) is reduced to a single Schr\"odinger equation with a scalar potential. In accordance with (\ref{c1}) this potential should be a rotational scalar, i.e., a function of $r^2=x_1^2 +x_2^2+...+x_d^2$. In addition, it has to  satisfy  one more condition, i.e.,  (\ref{c2}) (since (\ref{c3}) turns to identity). The general solution of (\ref{c2}) for $V=V(r)$ is:
\begin{gather}\la{kp}V=-\frac\alpha{r}\end{gather}
where $\alpha$ is a constant. The corresponding Schr\"odinger equation (\ref{se}) is superintegrable and admits a $d-$dimensional analogue of the LRL vector given by equation (\ref{lrl}), where \begin{gather}\la{lmn}J_{\mu\nu}\to L_{\mu\nu}=x_\mu p_\nu-x_\nu p_\mu\end{gather} and $V$ is the $d$-dimensional Coulomb potential (\ref{kp}).

Thus we recover a well known result \cite{Sud} concerning the generalization of the LRL vector in $d$ dimensions. Moreover, we also present a formal proof that the only scalar potential which is compatible with the $d$-dimensional LRL vector is the one given by equation (\ref{kp}).

In the following we restrict ourselves to attractive potentials with $\alpha>0$.

\subsection{Systems with spin $\frac12$}
Consider now a more complicated case when matrices $S_{\mu\nu}$ in (\ref{Jmn}) are non-trivial. The corresponding eigenvalue problem  (\ref{se}) includes a system of coupled Schr\"odinger equations in $d$ dimensional space.

Let us restrict ourselves to the cases when matrices $S_{\mu\nu}$ realize irreducible representation $D(\frac12,\frac12,...,\frac12)$ of algebra so($d$) for even $d$ or representation $D(\frac12,\frac12,...,-\frac12)$ for $d$ odd. Here the symbols in brackets are the Gelfand-Tsetlin numbers \cite{Gelf}.  Making reduction of these representations on subalgebra so(3) we obtain a direct sum of representations D($\frac12$). Thus it is possible to interpret the corresponding equations (\ref{se}) as a model of a particle with spin $\frac12$.

The considered matrices $S_{\mu\nu}$ admit the following uniform representation
\begin{gather}\la{a45}S_{\mu\nu}=\frac14\left(\gamma_\mu\gamma_\nu-
\gamma_\nu\gamma_\mu\right)\end{gather} where $\gamma_\mu$ are basis elements of the Clifford algebra  satisfying  the following relations
\begin{gather}\la{a5}\gamma_\mu\gamma_\nu+\gamma_\nu\gamma_\mu= 2\delta_{\mu\nu}.\end{gather}

The dimension of irreducible matrices (\ref{a45}) is equal to $2^{\left[\frac{d}2\right]}$ where $\left[\frac{d}2\right]$ is the entire part of $\frac{d}2$.

Let us search for potentials $V$ which satisfy relations (\ref{c1}), (\ref{c2}) and (\ref{c3}) together with matrices (\ref{a45}). The generic form of potential satisfying (\ref{c1}) is given by the following equations:
\begin{gather}\la{a6} V= f_1(r) +f_2(r)\gamma_\nu x_\nu
\end{gather}
for $d$ odd, and
\begin{gather}\la{a7} V= f_3(r)+f_4(r)\gamma_\nu x_\nu+f_5(r)\gamma_{d+1}\gamma_\nu x_\nu
\end{gather} for $d$ even.

Here $f_1(r), ..., f_5(r)$ are arbitrary functions.
Condition (\ref{c2}) specifies  these functions: $f_\nu=\frac{\alpha_\nu}{r^2}$ where $\nu=1, 2, ..., 5$ and $\alpha_\nu$ are constants.
The remaining condition (\ref{c3}) reduces potentials (\ref{a6}) and (\ref{a7}) to the following unified form:
\begin{gather}\la{a8}\hat V=\frac{\alpha}r\gamma_\nu n_\nu\end{gather}
where $n_\nu=\frac{x_\nu}r$.

Thus we specify a $d$-dimensional  system with spin $\frac12$ which is invariant with respect to group SO(d) and admits the generalized LRL vector. The corresponding  hamiltonian (\ref{se}) includes  potential (\ref{a8}), i.e., has the following form:
\begin{gather}\la{a9} H=\frac{p^2}{2m}+\frac{\alpha}{r^2}\gamma_\nu x_\nu.\end{gather}
For $d$=2 and $d=3$ this operator is equivalent to hamiltonians discussed in papers \cite{Pron} and \cite{N1} respectively.

The spectra of hamiltonians (\ref{a9}) and solutions of the corresponding equations (\ref{se}) are presented in section 3.
\subsection{Systems with spin 1}
Consider a bosonic $d$-dimensional system admitting generalized LRL vector (\ref{lrl}). We suppose the corresponding  matrices  $S_{\mu\nu}$ are irreducible and  realize representation D(1,0,0,...0) of algebra so($d$), where the symbols in brackets are the Gelfand-Tsetlin numbers. Up to equivalence, their entries $\left(S_{\mu\nu}\right)_{ab}$ can be represented in the following form:
\begin{gather}\la{s2}\left(S_{\mu\nu}\right)_{ab}=\ri(\delta_{\mu a}\delta_{\nu b}-\delta_{\nu a}\delta_{\mu b}),\quad \mu, \nu,\ a, b=1,2,...d.\end{gather}
Being reduced to its subalgebra so(3), algebra of matrices (\ref{s2}) is decomposed to a direct sum of representations $D(1)\oplus D(0)\oplus D(0)...$ which includes a spin-one (vector) representation $D(1)$ and $d-3$ scalar representations $D(0)$.

There are three basic scalars commuting with the corresponding total orbital momentum (\ref{Jmn}): functions of module of the $d$-dimensional radius-vector  $r$ multiplied by the unit matrix and such functions multiplied by the scalar matrix $S_{\mu\nu}x_\nu S_{\mu\lambda}x_\lambda$. Thus we can search for potentials of the following form:
\begin{gather}\la{v3}V=\frac1r\left(c_1 +c_2\frac{S_{\mu\nu}x_\nu S_{\mu\lambda}x_\lambda}{r^2}\right)\end{gather} where $c_1$ and $c_2$ are constants.

Such potentials automatically satisfy conditions (\ref{c1}) and (\ref{c2}). Substituting (\ref{v3}) and (\ref{s2}) into (\ref{c3}) we obtain the following solution:
\begin{gather}\la{v4}V=\frac\alpha{(d-2)r}\left({(d-1)(d-4)}+ {2}S_{\mu\nu}n_\nu S_{\mu\lambda}n_\lambda\right),\quad d\neq2.\end{gather}
Using realization (\ref{s2}) for matrices $S_{\mu\nu}$ it is possible to find entries $V_{\mu\nu}$ of matrix potential (\ref{v4}) in the following form:
\begin{gather} V_{\mu\nu}=\frac\alpha{2r}\left({(d-3)}\delta_{\mu\nu}+2n_\mu n_\nu\right). \la{v5}\end{gather}

It is interesting to note that, in contrast with representation (\ref{v4}), formula (\ref{v5}) is valid for the case $d=2$ also. Moreover, in the cases $d=2$ and $d=3$ potential  (\ref{v5}) is equivalent to potentials discussed in \cite{N5} and \cite{N1} respectively.

Potential (\ref{v5}) admits one more matrix representation alternative to (\ref{v4}). Indeed, using matrices which realize the representation D(1,0,0,...,0) of algebra so($d+1$), we can construct the following form:
\begin{gather}\la{v6}\hat V =\frac\alpha{2r}\left(d-3+2(S_{d+1 \mu}n_\mu)^2\right)=\begin{pmatrix}\frac{(d-1)\alpha}{2r}&0\\0^\dag&V
\end{pmatrix}\end{gather}where 0 is the $d\times 1$ zero matrix and $V$ is a $d\times d$ matrix whose entries are given by equation (\ref{v5}). In other words, it is a direct sum of the Coulomb potential and potential (\ref{v5}). For $d=2$ potential (\ref{v6}) is equivalent to potential for spin 1 discussed in \cite{Pron2}.

Hamiltonian (\ref{se}) with potentials (\ref{v4}) -- (\ref{v6}) commutes with the total orbital momentum (\ref{Jmn}) and the LRL vector (\ref{lrl}) where $S_{\mu\nu}$ are matrices (\ref{s2}). It means that the corresponding Schr\"odinger equation (\ref{se}) is possessed of Fock symmetry.
\section{Exact solutions}
\subsection{Solutions for scalar equations}
Thanks to their rotation invariance all equations presented above admit solutions in separated variables. For scalar equations such solutions are well known and they will be a starting point in our discussion.

To separate variables in equation (\ref{se}) with potential (\ref{kp})
it is possible to use the hyper-spherical variables which are related to the Cartesian variables via the following relations:
\begin{gather}\la{rv}\begin{split}&
x_d = r \cos \theta_{d-1},\\&
x_{d-1} = r \sin \theta_{d-1} \cos \theta_{d-2},\\&
x_{d-2} = r \sin \theta_{d-1} \sin \theta_{d-2} \cos \theta_{d-3},\\&
...\\&
x_2 = r \sin \theta_{d-1} \sin \theta_{d-2} . . . \sin \theta_{2} \cos \theta_{1},\\&
x_1 = r \sin \theta_{d-1} \sin \theta_{d-2} . . . \sin \theta_{2} \sin \theta_1.\end{split}
\end{gather}
In addition, wave function $\psi$ should be expressed via hyper-spherical harmonics $Y^l_\lambda$
\begin{gather}\la{hsh}\Psi=r^{\frac{1-d}2}
\psi_{l\lambda}Y^l_\lambda\end{gather}
where $Y^l_\lambda$ satisfy the following condition:
\begin{gather}\la{L2}\frac12L_{\mu\nu}L_{\mu\nu}Y^l_\lambda=
l(l+d-2)Y^l_\lambda,\end{gather}
and $\lambda$ is the multiindex enumerating eigenvalues of other Casimir operators of algebra so($d$) whose generators  $L_{\mu\nu}$ are given by equation (\ref{lmn}).

Substituting (\ref{hsh}) into (\ref{se}) we come to the following equation for the radial wave function:
\begin{gather}\la{re} H_\mu\psi_{l\lambda}(r)\equiv\left(-\frac{\p^2}{\p r^2}+\frac{\mu(\mu+1)}{r^2}+\hat V\right)\psi_{l\lambda}(r)=\epsilon\psi_{l\lambda}(r)\end{gather}
where $\hat V=2mV=-\frac{2m\alpha}r$, $\epsilon=2mE$ and
\begin{gather*}\mu=l+\frac{d-3}2, \quad l=0, 1, 2, ...\end{gather*}

Notice that admissible values of $E$ can be found algebraically, without solving equation (\ref{re}). Indeed, we have one more constraint, i.e., equation  (\ref{AK}).  For trivial matrices $S_{\mu\nu}$ this equation is reduced to the algebraic condition
\[\left(\frac12(d-1)^2mE+m^2\alpha^2\right)\Psi=-2mE\omega\Psi\]
provided equation (\ref{se}) is satisfied. Thus eigenvalues $E$ can be expressed via the spectral parameter $\omega=n(n+d-1), \ n=0, 1, 2,...$ whose values can be found within the representation theory of group so($d+1$) \cite{Kir}. Thus
\begin{gather}\la{spectr}E=-\frac{m\alpha^2}{2N^2}\end{gather}
where
\begin{gather}\la{N} N=n+l+\frac{d-1}2,\quad n=0, 1, 2, ...\end{gather}
For $d=3$ equation (\ref{spectr}) is reduced to the familiar Balmer formula.

Up to the meaning of quantum number $\mu$ equation (\ref{re}) coincides with the radial equation for the 3$d$ Hydrogen atom. Its solutions can be found using tools of supersymmetric quantum mechanics. Indeed, hamiltonian $H_\mu$ can be factorized:
\begin{gather}H_\mu=a^+_\mu a_\mu+c_\mu\la{fac}\end{gather}
where
\begin{gather}a_\mu=\frac{\p}{\p r}+W_\mu\la{amu}\end{gather}
with $\quad W_\mu=
-\frac{m\alpha}{\mu+1}+\frac{\mu+1}r$ and $c_\mu=\frac{(m\alpha)^2}{(\mu+1)^2}.$
 Moreover, the following intertwining relations are satisfied:
 \begin{gather}\la{ir}a^+_{\mu+1}H_{\mu+1}=H_\mu a_\mu^+.\end{gather}

 The ground state vector $\psi^0_\mu$ should solve the first order equation $a_\mu^+\psi^0_\mu=0$ and so has the following form
 \begin{gather}\la{gs}\psi^0_\mu=C_{0\mu}r^{\mu+1}\exp\left(\frac{-\alpha r m}{2(\mu+1)^2}\right).\end{gather}
 Then vectors of exited states $\psi^n_\mu$ and the corresponding eigenvalues $\epsilon_n$ are easily found using the following relation:
 \begin{gather}\label{es}\psi^n_{\mu}=
a_{\mu}^+a_{\mu+1}^+ \cdots
a_{\mu+n-1}^+\psi^0_{\mu+n,k},\quad\epsilon_n=-\frac{(m\alpha)^2}{4(\mu+n+1)^2}.
\end{gather}

In this way we obtain
\begin{gather}\la{phi1}\psi^n_\mu=C_{n\mu} z^{\mu+1}\exp(-z){\cal F}(-n,2\mu+2,2z)\end{gather} where $\cal F(.,.,.)$ is the confluent hypergeometric function, $z=\frac{m\alpha r}{n+\mu+1}$. The corresponding energy levels are  $E=-\frac{\epsilon_n}{2m}$, i.e., coincide with (\ref{spectr}).
\subsection{Solutions for spinor systems}
Consider now equations (\ref{se}) with matrix hamiltonian (\ref{a9}). To separate variables it is sufficient to use hyper-spherical variables (\ref{rv}) and  expand solutions as:
\begin{gather}\la{psi}\Psi({\bf r})=r^{\frac{d-1}2}
\psi_{j\varrho\lambda}(r)\Omega^j_{\varrho\lambda}\end{gather}
where $\Omega^j_{\varrho\lambda}$ are hyper-spherical spinors satisfying the conditions
\begin{gather}\begin{split}&\frac12J_{\mu\nu}J_{\mu\nu}
\Omega^j_{\varrho\lambda}=
\left(j(j+d-2)+\frac18(d-2)(d-3)\right)\Omega^j_{\varrho\lambda},
\\&D\Omega^j_{\varrho\lambda}\equiv\left(\frac12 \gamma_\mu\gamma_\nu L_{\mu\nu}+\frac{d-1}2\right)
\Omega^j_{\varrho\lambda}=\varrho\Omega^j_{\varrho\lambda}.\end{split}\la{casim}\end{gather}
Here $j$ and $\varrho$ are quantum numbers which take the following values
\[j=\frac12, \frac32, ...,\quad \varrho=\pm\left(j+\frac{d-2}2\right).\]

Substituting (\ref{a8}) and (\ref{psi}) into (\ref{se}) we come to the following equation for radial wave functions (see Appendix A):
\begin{gather}\la{req}H_\varrho\phi\equiv\left(-\frac{\p^2}{\p r^2}+\frac{\varrho^2+\sigma_3\varrho}{r^2}+
\frac\omega{r}\sigma_1\right)\phi=\varepsilon\phi,\qquad \phi=\begin{pmatrix}\psi_{j|\varrho|\lambda}(r)\\
\psi_{j-|\varrho|\lambda}(r)\end{pmatrix}\end{gather}
where $\varepsilon=2mE,\ \omega=2m\alpha$, $\sigma_1$ and $\sigma_3$ are Pauli matrices.

Equation (\ref{req}) includes effective potential $V_\varrho=\frac{\varrho^2+\sigma_3\varrho}{r^2}+
\frac\omega{r}\sigma_1$ which belongs to the list of shape invariant potentials presented in paper \cite{N3}, see equation (5.11) for $\mu=\varrho-\frac12$ and $\kappa=\frac12$ therein. The general  solutions of this equation for negative eigenvalues $E$ has been obtained in \cite{N3} using tools of SUSY quantum mechanics.

Let us apply the generic results presented in \cite{N3} to our particular system. First we note that Hamiltonian (\ref{req}) can be factorized, i.e., represented in the form (\ref{req}) where
\[a_\varrho=-\frac{\p}{\p \varrho}+W_\varrho,\quad W_\varrho=\frac{\sigma_3-2\varrho-1}{2r}+\frac{\omega\sigma_1}{2\varrho+1}.\]
The ground state vector  $\phi^0_\varrho$ is a solution of the first order equation $a\varrho^+\psi^0_\varrho=0$ whose explicit form is given by the following equation:
\begin{gather}\la{bst}\phi^0_\varrho=Cy^{\varrho+1}\begin{pmatrix}
K_1(y)\\K_0(y)\end{pmatrix}\end{gather}
where $K_0(y)$ and $K_1(y)$ are the modified Bessel functions, $y=\frac{\omega r}{2\varrho+1}$, and $C$ is the integration constant: \begin{gather*}C=2^{-2\varrho}\left(G^{22}_{00}\left(1|^{-1,0}_{\varrho-\frac12,
\varrho+\frac12}\right)+G^{22}_{00}\left(1|^{0,0}_{\varrho+\frac12,
\varrho+\frac12}\right)\right)^{-1}\end{gather*} with $G^{22}_{00}(.|^{..}_{..})$ being the Meijer functions.

The exited states vectors $\phi^n_\varrho$ are defined as
\begin{gather}\la{bst2}\phi^n_{\varrho}=
a_{\varrho}^+a_{\varrho+1}^+ \cdots
a_{\varrho+n-1}^+\psi^0_{\varrho+n}\end{gather}
while the corresponding  eigenvalues $E$ are given by equation (\ref{spectr}) where
\begin{gather}\la{NN}N=\varrho+n=j+n+\frac{d-1}2,\quad n=0,1,2,...\end{gather}

All vectors (\ref{bst}) and (\ref{bst2}) are square integrable and vanish at $r=0$.
In particular cases $d=2$ and $d=3$ they  are reduced to solutions discussed in \cite{Hau}, \cite{N5} and \cite{N1}.

Consider also the additional condition (\ref{AK}) which fixes the eigenvalue of the second order Casimir operator of the hidden symmetry algebra so($d$+1). Substituting
(\ref{a4}) and (\ref{a8}) into (\ref{AK}) and using the found eigenvalues $E$ we come to following algebraic relation
\begin{gather*} 2E(\omega+\frac{d(d-1)}8+\alpha^2m=0\end{gather*}
or
\begin{gather}\omega={\hat j}({\hat j}+d-1)+\frac{(d-1)(d-2)}8\la{ome}\end{gather}
where $\hat j=j+n=\frac12, \frac32,...$ Notice that eigenvalues (\ref{ome}) which correspond to algebra so($d$+1) and eigenvalues in the first line of equation (\ref{casim}) which correspond to algebra so($d$) are connected via the excepted relation $d\to d-1$.
\subsection{Solutions for vector systems}
Finally, let us discuss equation (\ref{se}) with matrix potential whose entries are given by equation (\ref{v5}). Solutions of this equation are $d$-component vectors $\Psi=$column$(\Psi_1, \Psi_2, ..., \Psi_d)$ which form the space of irreducible representation $D(1,0,0,...,0)$ of group SO($d$).

We will consider (\ref{se}) together with the supplementary condition
(\ref{AK}). Using equation (\ref{se}) and the following identities
\begin{gather*}\begin{split}&S_{\mu\nu}L_{\mu\nu}V+VS_{\mu\nu}L_{\mu\nu}=
\frac{\alpha}r\left((d-2)S_{\mu\nu}L_{\mu\nu}+1+
\frac{d(d-3)}2\right)+(d-2)V,\\
&\frac12S_{\mu\nu}S_{\mu\nu}=d-1,\quad S_{\lambda\mu}S_{\lambda\nu}p_\mu p_\nu=p^2+(d-2)P\end{split}\end{gather*}
where $P$ is a matrix with entries $P_{\mu\nu}=p_\mu p_\nu$,
 we reduce (\ref{AK}) to the following form:
\begin{gather}\la{AC}\begin{split}&\left(p_\mu p_\nu-\frac{\alpha m}x(L_{\mu \nu}+\delta_{\mu\nu}-n_\mu n_\nu)+m^2\alpha^2n_\mu n_\nu+\varepsilon\right)\Psi_\nu=0\end{split}\end{gather}
where
\begin{gather}\la{var}\varepsilon=\frac1{4(d-2)}\left(2mE
\left({(d-3)^2}+
4\omega\right)+{m^2\alpha^2(d-3)^2}\right). \end{gather}

Equations (\ref{se}) and (\ref{AC}) where $V$ is potential (\ref{v5}) admit solutions in separated variables.
To separate variables we
represent entries of the  wave function as linear combinations of the following linearly independent terms:
\begin{gather}\la{vh3}\Psi_\mu=\varphi^1_{l\lambda}(r)\Phi^1_\mu +\varphi^2_{l\lambda}(r)\Phi^2_\mu+\Phi^3_\mu\end{gather}
where
\begin{gather}\la{vh}\Psi^1_\mu=\frac{x_\mu}r Y^l_\lambda,\quad \Psi^2_\mu =r\nabla_\mu Y^l_\lambda,\end{gather}
$Y^l_\lambda$ are hyper-spherical harmonics discussed in section 3.1, and $\Phi^3_\mu$ is a vector satisfying the following relations:
\begin{gather}\la{vh2} x_\mu\Phi^3_\mu=0,\quad \quad \nabla_\mu\Phi^3_\mu=0.\end{gather}

For $d=3$ vector $\Phi^3_\mu$ can be represented as $\Phi^3_\mu=\frac12\varepsilon_{\mu\nu\sigma}L_{\nu\sigma}
\varphi^3_{l\lambda}(r)Y^l_\lambda$ were $\varepsilon_{\mu\nu\sigma}$ is the Levi-Civita tensor. In this case $\Psi^1_\mu, \Psi^2_\mu$ and $\Psi^1_\mu$ are nothing but the familiar vector harmonics.
We will not specify $\Phi^3_\mu$ for arbitrary $d$ since it will not be present in the final solutions.

Introducing hyper-spherical variables (\ref{rv}) and
substituting (\ref{v5}) and (\ref{vh3}) into (\ref{se}) we recognize that the latter system is decoupled to two subsystems. One of them involves only $\Phi^3_\mu$ and has the following form:
\begin{gather}\la{ph3}\left(p^2+\frac{\tilde\alpha}r\right)\Phi^3_\mu
=2mE\Phi^3_\mu\end{gather}
where $\tilde\alpha=m(d-3)\alpha$. The other subsystem includes equations for radial functions $\varphi_1=\varphi_1^{l\lambda}$ and $\varphi_2=\varphi_2^{l\lambda}$ and can be written as:
\begin{gather}\la{vph1}\begin{split}&-\varphi_1''-\frac{d-1}r\varphi_1'+
\frac1{r^2}((l(l+d-2)+d-1)\varphi_1-2l(l+d-2))\varphi_2+
\frac{m\alpha(d-1)}r\varphi_1\\&=2mE\varphi_1,\end{split}\\\la{vph2}
-\varphi_2''-\frac{d-1}r\varphi_1'+
\frac1{r^2}((l(l+d-2)-d+3)\varphi_2-2\varphi_1)+
\frac{m\alpha(d-3)}r\varphi_2=2mE\varphi_2\end{gather}
where $\varphi_1'=\frac{\p \varphi_1}{\p r}$, etc.

The system (\ref{AC}) can be decoupled too. Namely, substituting (\ref{vh3}) into (\ref{AC}) and using (\ref{vph1}) we obtain:
\begin{gather}\varepsilon\Phi_\mu=0,\la{AC3}\\\la{AC1}
l(l+d-2)(\varphi_1-(r\varphi_2)'+\alpha mr\varphi_2)-r^2(m^2\alpha^2+2mE+\varepsilon)\varphi_1=0,\\
(d-2){\varphi_1}+{(r\varphi_1)'}-l(l+d-2)
{\varphi_2}+{m\alpha}r\left(\varphi_1-
\frac{d-3}2\varphi_2\right)-r^2\varepsilon\varphi_2=0.
\la{AC2}
\end{gather}

Thus we have two separated systems of equations. The first of them describes functions $\Phi^3_\mu$ and includes equations (\ref{ph3}) and (\ref{AC3}). The other one involves equations (\ref{vph1}), (\ref{vph2}) and (\ref{AC1}), (\ref{AC2}) for variables $\varphi_1$ and $\varphi_2$.

Let us start with the latter equations, i.e., the first order system (\ref{AC1}), (\ref{AC2}) and the second order system (\ref{vph1}), (\ref{vph2}). The compatibility condition of these systems  reads:
\begin{gather}\la{CC}2mE+\varepsilon+\alpha^2m^2=0.
 \end{gather}
Indeed, differentiating equations (\ref{AC1}) and (\ref{AC2}) and expressing the second order derivatives in accordance with (\ref{vph1}) and (\ref{vph2}) we immediately come to (\ref{CC}), otherwise   $\varphi_1$ and $\varphi_2$ should be trivial.

Equations (\ref{CC}) and (\ref{var}) make it possible to express  energy levels $E$ via eigenvalues $\omega$ of the second order Casimir operator of algebra so($d+1$):
\begin{gather}\la{E1} E=-\frac{m\alpha^2(d-1)^2}{2(d-1)^2+8\omega}. \end{gather}

The last term in the left hand side of equation (\ref{AC1}) is equal to zero, therefore
\begin{gather}\la{AC4}\varphi_1=(r\varphi_2)'-\alpha m r\varphi_2.\end{gather}
Substituting that into (\ref{vph2}) and setting $\varphi_2=r^{-\frac{d+1}2}\phi$ we obtain the following equation:
\begin{gather}-\phi''+\left(\frac{\mu(\mu+1)}{r^2}-
\frac\kappa{r}\right)\phi=2mE\phi\la{se2}\end{gather}
where $\kappa=\alpha(d-1)m$ and $\mu=l+\frac{d-3}2$. In other words, we again recognize equation (\ref{re}) where, however,  $\alpha\to\frac12(d-1)\alpha$. Thus the admissible eigenvalues $E$  and the corresponding solutions $\phi$ of (\ref{se2}) are given by formulas (\ref{spectr}) and (\ref{phi1}) with $\alpha\to\frac12(d-1)\alpha$. The related functions $\varphi_1$ are easy calculated using equation (\ref{AC4}). Thus
\begin{gather}\la{spectra}
E=-\frac{m\alpha^2}{2k^2}\end{gather}
where $k=\frac{2n+2l+d-1}{d-1}$.

Comparing energy levels (\ref{spectra}) and (\ref{E1}) we obtain the spectrum of the second order Casimir operator $C_2$:
\[\omega=l'(l'+d-1) \quad \text{where} \quad l'=l+n=0, 1, 2, ...\]

The radial wave functions are:
\begin{gather}\la{ei1}\begin{split}&\varphi_2=C_{ln}z^{l+d-1}\exp(-z){\cal F}(-n,l+d-1,2z),\\&\varphi_1=C_{ln}\left(z^{l+d-1}\exp(-z)\left((l+d+(k-1)z){\cal F}(-n,l+d-1,2z)\right.\right.\\&\left.\left.-\frac{2kz}{l+d-1}{\cal F}(-n+1,l+d,2z)\right)\right)\end{split}
\end{gather}
where $z=\frac{m\alpha r}{k}.$ They
are normalizable and vanish at $z=0$.

Consider now equations (\ref{ph3}) and (\ref{AC3}). The first of them fixes admissible energy values in the following form (see definition (\ref{var})):
\begin{gather}\la{E2} E=-\frac{m\alpha^2(d-3)^2}{2(d-3)^2+8\omega}. \end{gather}
This relation is incompatible with (\ref{E1}). Thus solutions with non-trivial $\Psi_\mu^3$ correspond to trivial $\varphi_1$ and $\varphi_2$ and vise versa. It happens that in fact $\Psi_\mu^3$ should be trivial (see Appendix B) while $\varphi_1$ and $\varphi_2$ are presented by equations (\ref{ei1}).

\section{Discussion} Thus we extend the field of superintegrable systems admitting LRL vector. Originally this vector (both classical and quantum mechanical) was specified for three dimensional space. Then it was generalized to the case of $d$ dimensions \cite{Sud}, but it was done only for scalar QM systems.

 The first example of LRL vector with spin was proposed apparently in paper \cite{Pron}, where a 2$d$ superintegrable system was discovered.
LRL vectors with spin in three dimensional space were discussed in \cite{N1} and \cite{N6}.

In the present paper the results of papers
\cite{N1} and \cite{N6} are extended to the case of arbitrary dimension. Effectively, the number of presented systems is infinite, but countable. All of them are exactly solvable, and this property is used to obtain their exact solutions in explicit forms.

We restrict ourselves to discussion of systems with spins 0, 1/2 and 1, which correspond to the trivial representation of algebra so($d$) and also irreducible representations $D(\frac12,\frac12,...,\frac12)$, $D(\frac12,\frac12,...,-\frac12)$ and $D(1,0, ...,0)$ of this algebra. Starting with other representations and using the determining equations (\ref{c1}), (\ref{c2}) and (\ref{c2}) it is possible to construct additional superintegrable systems with spin, admitting the LRL vector. However, in contrast with the 2d and 3d cases \cite{N5}, \cite{N6}, it is seemed to be impossible to construct explicitly  the  generic model for arbitrary irreducible representation of algebra so($d$).

The $d$ dimensional  systems with spin admitting LRL keep all
basic properties of the systems with $d$=3. First, they are superintegrable. Their extended symmetries present effective tools for finding exact solutions. In particular, these symmetries make it possible to impose the additional condition (\ref{AK}) on solutions of the Schr\"odinger equation. For systems with spin 0 and $\frac12$ this condition is reduced to algebraic equations for the Hamiltonians eigenvalues while for the case of spin one we have a system of first order equations  (\ref{AC1}) and (\ref{AC2}).

The very existence of the first order system compatible with the initial (second order) equations can be interpreted as a conditional symmetry, see \cite{Rena} for exact definitions. But in contrast with the generic conditional symmetry which gives rise to existence of particular exact solutions, its particular case generated by the LRL vector makes it possible to find the {\it general} solution of equation (\ref{se}).

The discussed systems with spin $\frac12$ are supersymmetric, i.e., the corresponding radial equations are shape invariant. Moreover, they belong to the list of shape invariant systems classified in \cite{N3} and \cite{N4}.  Effectively, in the present paper  a countable set of systems  with shape invariant matrix potentials is constructed. As it was shown in Section 3.2, all these systems  can be successfully  solved using tools of SUSY quantum mechanics. The same is true for the scalar systems whose general solutions are presented in section 3.1.

The considered systems with spin 1 are neither supersymmetric nor shape invariant. However, they are superintegrable and exactly solvable, see section 3.3 for the explicit solutions. Supersymmetric systems in $d$ dimensions were discussed in paper \cite{Kir}.

Some results of the present paper had been announced in the conference proceedings published in \cite{NNN}.

\section{Appendix A. Separation of variables in spinor equations}
\renewcommand{\theequation}{A\arabic{equation}} %
\setcounter{equation}{0}
A standard way to separate variables in equation invariant w.r.t. algebra so($d$) is to expand solutions via eigenvectors of the complete set of $d-1$ commuting integrals of motion, which includes the Casimir operators of algebra so($d$) (whose generators are given by equations (\ref{Jmn}) and (\ref{a4})) and also operators $J_{12}, J_{34}, ...$ and change Cartesian coordinates to radial and hyper-spherical variables. Let present a  more simple and straightforward way to obtain the radial equations discussed in section 3.2.

  First we define the following operator
\begin{gather} D=\frac12 \gamma_\mu\gamma_\nu L_{\mu\nu}+\frac{d-1}2.\la{di}\end{gather}
Its important properties are:
\begin{itemize}
\item $D$ anticommutes with operators $\gamma_\mu p_\mu$ and  $\gamma_\mu x_\mu$
\begin{gather}\la{ak}D\gamma_\mu p_\mu=-\gamma_\mu p_\mu D,\quad \gamma_\mu x_\mu=-\gamma_\mu x_\mu D;\end{gather}
\item In a space of eigenvectors of the second order Casimir operator $C_2=\frac12J_{\mu\nu}J_{\mu\nu}$ the square of $D_d$ is proportional to the unit matrix, since
    \begin{gather}\la{dsr}(D)^2=\frac12J_{\mu\nu}J_{\mu\nu}+
    \frac18(d-1)(d-2);\end{gather}
    Moreover, eigenvalues of operators $D$ and $C_2$ are given by formulae (\ref{casim}).
\item If $d$ is even then multiplying $D$ by  $\gamma_{d+1}=\frac1{d!}\varepsilon_{\mu_1\mu_2...\mu_d}\gamma_{\mu_1}\gamma_{\mu_2}... \gamma_{\mu_d}$ where $\varepsilon_{\mu_1\mu_2...\mu_d}$ is the absolutely antisymmetric unit tensor, we obtain an integral of motion for hamiltonian (\ref{a9}). For $d=3$ operator $\gamma_{d+1}D$ is reduced to Dirac constant of motion for the relativistic Hydrogen atom.
\end{itemize}

To deduce the equation for radial functions we use the identities \begin{gather}\la{p2}p^2=\left(-\ri\gamma_\mu \p_\mu\right)^2,\quad (-\ri\gamma_\nu n_\nu)^2=1, \quad \gamma_\mu\gamma_\nu=-\delta_{\mu\nu}-2\ri S_{\mu\nu}\end{gather}
 and evaluate $\gamma_\mu \p_\mu$ in the following way:
\begin{gather}\la{gmpm}\begin{split}&\gamma_\mu \p_\mu\equiv (-\ri\gamma_\nu n_\nu)^2\gamma_\mu \p_\mu=
-\ri\gamma_\nu n_\nu\left(n_\mu p_\mu+\ri\frac{d-1}{2r}-\frac{i}r D\right)\\&=-\ri\gamma_\nu n_\nu\left(-i\frac{\p}{\p r}+\ri\frac{d-1}{2r}-\frac{i}r D\right).\end{split}\end{gather}
Substituting (\ref{gmpm}) into (\ref{p2}) and using (\ref{ak}) we obtain the following
 convenient representation of the $d$ dimensional Laplace operator:
\begin{gather}\la{la}p^2=-\Delta_d=-\frac{\p^2}{\p r^2}-\frac{d-1}r\frac{\p}{\p r}-\frac1{4r^2}(d-1)(d-3)+\frac1{r^2}D(D+1)\end{gather}
which generates equation (\ref{req}) for radial wave function.

\section {Appendix B. Some calculation details for vector systems}
\renewcommand{\theequation}{B\arabic{equation}} %
\setcounter{equation}{0}
Considering equations  (\ref{ph3}) for wave functions $\Psi_\mu^3$ we conclude that this system  is completely decoupled. Moreover, for any fixed $\mu$ we have an equation which, up to the change $\alpha\to\tilde\alpha=\frac12(d-3)\alpha$, coincides with  the scalar equation solved in section 3.1. Thus we know both eigenvectors (which will be discussed later) and the corresponding eigenvalues:
\begin{gather}\la{E3}E=-\frac{m\alpha^2(d-3)^2}
{2\left(2l+2n+d-1\right)^2}.
\end{gather}

The compatibility condition for equations (\ref{E2}) and (\ref{E3}) can be written as
\begin{gather}\la{om}\omega=\tilde l(\tilde l+\tilde d-2)\end{gather}
where $\tilde l=l+n+1=1,2,..., \ \tilde d=d-1.$ Let us show that this relation is not realizable.

Equation (\ref{om}) should define the spectrum of the second order Casimir operator of algebra so($d+1$). This algebra includes subalgebra so($d$) whose basis elements are given by equations (\ref{Jmn}) and (\ref{s2}). The spectrum of the corresponding second order Casimir operator (\ref{acon}) with $\mu, \nu=1,2,...,d$ is given by the following relation:
\begin{gather}\la{sp}\omega=j(j+d-2),\quad j=0, 1, 2, ...\end{gather}
However, spectra (\ref{om}) and (\ref{sp}) are not compatible. More exactly, representations of subalgebra so($d$) which correspond to spectrum (\ref{sp}) cannot be obtained by reduction of hypothetical representations of algebra so($d+1$) corresponding to spectrum (\ref{sp}), since such reduction should be attended by {\it decreasing} of quantum number $d$, while in our case it {\it increases}. Thus eigenvalues (\ref{om}) for the Casimir operator and the corresponding eigenvalues (\ref{E3}) for the Hamiltonian are forbidden, and so vectors $\Phi^3_\mu$ in expansion (\ref{vh3}) should be trivial.

\end{document}